\documentclass[preprint,showpacs,preprintnumbers,amsmath,amssymb,nofootinbib]{revtex4}

\usepackage{graphicx}
\usepackage{dcolumn}
\usepackage{bm}

\begin{document}


\title{Consistency of particle-particle random-phase
approximation\\ and its renormalizations\footnote{to appear in
Physical Review C}}

\author{Nguyen Dinh Dang}
\altaffiliation[E-mail address: ]
{dang@riken.jp}
\affiliation{
1) Heavy-ion nuclear physics laboratory, Nishina Center for
Accelerator-Based Science, RIKEN, 2-1 Hirosawa, Wako city, 351-0198 
 Saitama, Japan
 \\
2) Institute for Nuclear Science and Technique, Hanoi, Vietnam}

\begin{abstract}
The consistency condition is tested within the 
particle-particle random-phase approximation (RPA), renormalized
RPA (RRPA) and the self-consistent RPA (SCRPA) making use of the
Richardson model of pairing. The two-particle separation energy is
calculated in two ways, namely as the energy of  
the first addition mode, which adds two particles to a core with $N$ particles, 
and as the energy of the first removal 
mode, which removes two particles from the $N+2$ particle system to
get back to the same $N$-particle core.
The corresponding transitions generated by the pairing
operators are also calculated. It is found that the results obtained in
these two ways of calculations are close to each other only at large
values of particle number $N$ and/or small interaction strength. 
At $N\leq$ 10 for a given value of
the interaction strength, 
the discrepancy between the results obtained in two ways of
calculations within the SCRPA is much smaller than those given by the RPA 
and RRPA. 
\end{abstract}
\maketitle
\section{Introduction}
The random-phase approximation (RPA) has been widely used
in the theoretical study of nuclei within the valley of $\beta$-stability.
The success of the RPA is mainly based on the use of the quasiboson 
approximation (QBA), which considers fermion pairs as boson
operators, just neglecting the Pauli principle between them.
As a result, a set of linear equations, called the RPA equations, 
was derived, which reveals the physics of collective excitations 
generated by the RPA boson-like modes. The simplicity 
of the RPA equations allows a feasible treatment of
a number of many-body problems, which  would be 
computationally intractable otherwise. However, this approach 
suffers a drawback: It breaks down 
at a certain critical value of the interaction's parameter, 
where the RPA yields imaginary solutions. The reason of this 
well-known RPA instability is the violation of 
Pauli principle within the QBA. 

In $\beta$-stable medium and heavy nuclei, the QBA is a good 
approximation, 
and the RPA is a very powerful tool for the description of several 
important quantities such as the ground-state and 
excited-state energies,
electro-magnetic transition probabilities and their distribution, 
transition densities, etc. The first-order diagram 
expansion beyond the mean field treated within the RPA includes 
significant effects of two-body correlations beyond the mean-field. 
However, with reducing the particle number, 
the concept of collective excitations, which are described by the RPA 
modes, becomes less and less firm.
The ground-state correlations (GSC) which are left beyond the 
RPA become stronger in light systems. This feature makes
the validity of the QBA, and therefore of the RPA itself, questionable
in the systems with small particle numbers.

Several approaches were developed to take into account the GSC beyond 
RPA in a simple way such as to restore the Pauli principle among 
the fermion pairs, from which the RPA operators are constructed. 
The popular one, known as the renormalized RPA 
(RRPA)~\cite{Hara,Rowe,Schuck,CaDaSa}, includes the expectation 
value over the ground state of the diagonal
elements of the commutator between two fermion-pair operators, 
neglected in the QBA. In this way the Pauli principle 
is approximately taken care of. 
The inclusion of GSC beyond RPA within the RRPA 
eventually renormalizes the interaction 
in such a way that the collapse of 
RPA is avoided, and the RRPA has solutions at any value of the 
interaction's parameter. This scheme has been tested in several exactly 
solvable schematic models such as the Lipkin model for $ph$
RRPA~\cite{Hara,CaDaSa}, the 
SO(8) model for the renormalized quasiparticle 
RPA~\cite{RQRPA}, and the Richardson model 
for the $pp$ RRPA and self-consistent RPA (SCRPA)~\cite{Duk,Dang}. 
The results of these studies showed that the 
difference between the energies of the first excited state given by 
the $pp$ RPA, $pp$ RRPA, and SCRPA increases noticeably with increasing the 
interaction parameter and/or decreasing the particle number. 
The SCRPA solutions are the closest ones to the exact solutions.
These results raise a serious question of the applicability of 
the RPA  to the calculations for 
realistic light-neutron rich nuclei. As a matter of fact, a close 
agreement between the RPA results and the experimental data for 
light neutron-rich nuclei can
be considered really reasonable if the corrections due to GSC beyond 
RPA are negligible, which justify the validity of the
QBA~\cite{VinhMau}. In the 
opposite case such agreement may be even illusory.

This concern is the motivation of the present work, whose goal is
to test the consistency of the $pp$ RPA and its renormalizations,
the $pp$ RRPA and SCRPA, at different particle numbers as the
interaction strength varies. The reason of choosing the $pp$ RPA comes
from one of its major merits, which allows the straightforward
calculation of an important quantity in the study of unstable nuclei, 
namely the two-particle separation energy. Indeed, the $pp$ RPA uses the addition 
and removal phonon operators to create the ground states and excited states of 
the systems with $N+2$ and $N-2$ particles, respectively, from the
ground state of the $N$-particle system. Therefore the two-particle
separation energy of the $(N+2)$-particle system can be calculated either as
the energy of the lowest excited state generated by the addition mode, 
which adds two particles to the core with N particles, or as that generated by the
removal mode, which removes two particles from the core with $N+2$
particles. A consistent theory should give identical results in either
way of calculation. By analyzing the results obtained in two ways of
calculations within all three approaches, namely the RPA, RRPA, and
SCRPA in their application to an exactly solvable schematic model,
namely the Richardson model~\cite{Ric}, the present work may shed
light on the consistency of
these approaches, in particular, in their application to 
systems with small particle numbers (light systems). 

The paper is organized as follows. The brief outline of the $pp$ RPA, 
RRPA, and SCRPA applied to the Richardson model is presented in Sec. \ref{outline}.
As the present paper deals only with $pp$ RPA and its renormalization,
the prefix $pp$ will be omitted hereafter.
The results of numerical 
calculations are analyzed in Sec. \ref{numerical}. The paper is 
summarized in the last section, where conclusions are drawn.
\section{RPA, RRPA and SCRPA within the Richardson model}
\label{outline}
\subsection{The Richardson model}
The Richardson model, considered in the present paper, consists of $\Omega$
doubly-fold equidistant levels, which interact via a pairing force with parameter $G$.  
The model Hamiltonian is given as 
~\cite{Duk,Dang} 
\begin{equation}
H=\sum_{i=1}^{\Omega}(\epsilon_{i}-\lambda)N_{i}-G\sum_{i,j=1}^{\Omega}
P_{i}^{\dagger}P_{j}~,
\label{H}
\end{equation}
where the particle-number operator $N_{i}$ and pairing operators 
$P_{i}^{\dagger}$, $P_{i}$ are defined as
\begin{equation}
N_{i}=c_{i}^{\dagger}c_{i}+c_{-i}^{\dagger}c_{-i}~, 
\hspace{5mm} 
P_{i}^{\dagger}=c_{i}^{\dagger}c_{-i}^{\dagger}~,\hspace{5mm} 
P_{i}=(P_{i}^{\dagger})^{\dagger}~.
\label{N&P}
\end{equation}
These operators fulfill the following exact commutation relations
\begin{equation}
[P_{i}, P_{j}^{\dagger}]=\delta_{ij}(1-N_{i})~,
\label{[PP]}
\end{equation}
\begin{equation}
[N_{i}, P_{j}^{\dagger}]=2\delta_{ij}P_{j}^{\dagger},~\hspace{5mm} [N_{i}, 
P_{j}]=-2\delta_{ij}P_{j}~.
\label{[NP]}
\end{equation}
The single-particle energies of the equidistant levels are defined as 
$\epsilon_{i}=i\epsilon$ with $i$ running over all $\Omega$ levels.
The present paper considers only the 
$ph$-symmetric case. This means that,
in the absence of interaction ($G=$0), the lowest $\Omega/2$ levels are filled
with $N=\Omega$ particles (two particles on each level).
Numerating particle ($p$) and hole ($h$) levels 
from the levels nearest to the Fermi level, 
the particle and hole energies are equal to 
$\epsilon_{p}=\epsilon(\Omega/2+p)$ and 
$\epsilon_{h}=\epsilon(\Omega/2-h+1)$, 
respectively, with 
$p$ ($h$) $=$ 1, \ldots, $\Omega/2$.
The Fermi level $\epsilon_{\rm F}$  
is defined as 
\begin{equation}
    \epsilon_{\rm F} =[\epsilon(\Omega+1)-{G}]/2~.
    \label{lambda}
\end{equation}
There are several methods of solving exactly 
the pairing problem described by Hamiltonian (\ref{H}), namely the 
Richardson method~\cite{Ric}, the infinite-dimensional algebras in Ref. 
\cite{Pan}, and the direct diagonalization of 
Hamiltonian (\ref{H}) in the Fock space~\cite{Volya}.  
\subsection{RPA}
The outline of the RPA is given in detail in Ref. \cite{Ring}. 
It uses the addition 
$A^{\dagger}_{\mu}$ and removal $R^{\dagger}_{\lambda}$ 
phonon operators to create the ground states and excited states of 
the systems with $N+2$ and $N-2$ particles, respectively, from the
ground state of the $N$-particle system. Applied to the Richardson
model, these addition and removal 
phonon operators are given as
\begin{equation}
{A}_{\mu}^{\dagger}=\sum_{p}
X_{p}^{(\mu)}Q^{\dagger}_{p}
-\sum_{h}Y_{h}^{(\mu)}
Q_{h}~,
\label{A}
\end{equation}
and
\begin{equation}
{R}_{\lambda}^{\dagger}=\sum_{h}
X_{h}^{(\lambda)}Q^{\dagger}_{h}
-\sum_{p}
Y_{p}^{(\lambda)}Q_{p}~,
\label{R}
\end{equation}
respectively, where
\begin{equation}
Q^{\dagger}_{p}=P^{\dagger}_{p}~,\hspace{5mm} Q_{h}=-P^{\dagger}_{h}~.
\label{Q}
\end{equation}
The RPA assumes the validity of the QBA, which replaces the
expectation value of the commutation relations
$[Q_{i},Q^{\dagger}_{j}]$ 
in the RPA ground state $|N,0\rangle$ of $N$-particle
system with that obtained in the
Hartree-Fock (HF) ground state $|{\rm HF}\rangle$, namely
       \begin{equation}
       \langle{N,0}|
	      [{Q}_{i}, {Q}_{j}^{\dagger}]|
	      {N,0}\rangle\simeq
	      \langle{\rm HF}|
			    [{Q}_{i}, {Q}_{j}^{\dagger}]|
			    {\rm HF}\rangle =
	      \delta_{ij}~,\hspace{5mm} (i,j)=(p,p')~, (h,h')~,
	      \label{QBA}
\end{equation}
since $\langle{\rm HF}|N_{p}|{\rm HF}\rangle=$ 0, and $\langle{\rm
HF}|N_{h}|{\rm HF}\rangle=$ 2.
Using the QBA (\ref{QBA}), 
one can easily verify that the expectation values of the commutation
relations for addition and removal operators in the RPA ground state
are of bosonic type, i.e.
\begin{equation}
    \langle N,0|[A_{\mu},A_{\mu'}^{\dagger}]|N,0\rangle=
    \delta_{\mu\mu'}~,\hspace{5mm} 
    \langle N,0|[R_{\lambda},R_{\lambda'}^{\dagger}]|N,0\rangle=
    \delta_{\lambda\lambda'}~,
    \label{[AA][RR]}
\end{equation}
if the $X$ and $Y$ amplitudes obey the following normalization 
(orthogonality) conditions 
\[
\sum_{p}X_{p}^{(\mu)}X_{p}^{(\mu')}-\sum_{h}Y_{h}^{(\mu)}Y_{h}^{(\mu')}=\delta_{\mu\mu'}~,
\hspace{2mm} 
\sum_{h}X_{h}^{(\lambda)}X_{h}^{(\lambda')}-\sum_{p}Y_{p}^{(\lambda)}Y_{p}^{(\lambda')}
    =\delta_{\lambda\lambda'}~,\]
    \begin{equation}
    \sum_{p}X_{p}^{(\mu)}Y_{p}^{(\lambda)}-\sum_{h}X_{h}^{(\lambda)}Y_{h}^{(\mu)}=0~,
    \label{norm}
\end{equation}
while the closure relations 
\[
\sum_{\mu}X_{p}^{(\mu)}X_{p'}^{(\mu)}-\sum_{\lambda}Y_{p}^{(\lambda)}Y_{p'}^{(\lambda)}
=\delta_{pp'}~,\hspace{2mm} 
    \sum_{\lambda}X_{h}^{(\lambda)}X_{h'}^{(\lambda)}-\sum_{\mu}
    Y_{h}^{(\mu)}Y_{h'}^{(\mu)}
    =\delta_{hh'}~,\]
    \begin{equation}
    \sum_{\lambda}X_{h}^{(\lambda)}Y_{p}^{(\lambda)}-\sum_{\mu}X_{p}^{(\mu)}Y_{h}^{(\mu)}=0~
    \label{closure}
\end{equation}
guarantee the following inverse transformation of Eqs. (\ref{A}) and (\ref{R})
\begin{equation}
Q_{p}^{\dagger}=\bigg[\sum_{\mu}X_{p}^{(\mu)}A_{\mu}^{\dagger}
+\sum_{\lambda}Y_{p}^{(\lambda)}R_{\lambda}\bigg]~,\hspace{5mm} Q_{h}=\bigg[\sum_{\lambda}X_{h}^{(\lambda)}R_{\lambda}
+\sum_{\mu}Y_{h}^{(\mu)}A_{\mu}^{\dagger}\bigg]~.
\label{inverse}
\end{equation}
The RPA equations are obtained by linearizing the 
equation of motion with Hamiltonian (\ref{H}) and operators
$A^{\dagger}$ and
$R^{\dagger}$. The matrix form of the RPA equation for the 
addition mode is
\begin{equation}
\left(\begin{array}{cc}{A}&{B}\\
    -{B}&{C}\end{array}
    \right)\left(\begin{array}{c}{X^{(\mu)}}\\
    {Y^{(\mu)}}\end{array} \right)=
    E_{\mu}\left(\begin{array}{c}{X^{(\mu)}}\\
    {Y^{(\mu)}}\end{array}\right)~,
    \label{RPA}
\end{equation}
where the submatrices $A$, $B$, and $C$ are found
as~\cite{Duk,Ring}
\begin{equation}
 A_{pp'}^{\rm RPA}= \langle{\rm HF}|[{Q}_{p},
[H,{Q}_{p'}^{\dagger}]]|{\rm HF}\rangle
=2\bigg[\epsilon\bigg(p-\frac{1}{2}\bigg)+\frac{G}{2}\bigg]
 \delta_{pp'}-G~,
\label{ARPA}
\end{equation}
\begin{equation}
B_{ph}^{\rm RPA}=-\langle{\rm HF}|[{Q}_{p},
[H,{Q}_{h}]]|{\rm HF}\rangle=G~,
\label{BRPA}
\end{equation}
\begin{equation}
 C_{hh'}^{\rm RPA}=-\langle{\rm HF}|[{Q}_{h},
[H,{Q}_{h'}^{\dagger}]]|{\rm HF}\rangle=
-2\bigg[\epsilon\bigg(h-\frac{1}{2}\bigg)+\frac{G}{2}\bigg]\delta_{hh'}
+G~.
\label{CRPA}
\end{equation}
\subsection{RRPA}
The QBA (\ref{QBA}) neglects GSC 
beyond RPA. The latter make the expectation value of $N_{p}$ deviate from zero and 
that of $N_{h}$
deviate from 2.
In order to take this effect into account, the 
RRPA considers the following renormalized addition 
and removal phonon operators
\begin{equation}
    {\cal A}_{\mu}^{\dagger}=
    \sum_{p}{{X}_{p}^{(\mu)}}
    \overline{Q}_{p}^{\dagger}-
    \sum_{h}{{Y}_{h}^{(\mu)}}\overline{Q}_{h}~,    
    \hspace{5mm} {\rm and}\hspace{5mm}  
    {\cal R}_{\lambda}^{\dagger}=
    \sum_{h}
    {{ X}_{h}^{(\lambda)}}\overline{Q}_{h}^{\dagger}-
    \sum_{p}{Y}_{p}^{(\lambda)}\overline{Q}_{p}~,
\label{calAR}
\end{equation}
respectively~, with the abbreviation
\begin{equation}
\overline{\cal O}_{i}^{\dagger}=\frac{{\cal
O}_{i}^{\dagger}}{\sqrt{\langle D_{i}\rangle}}~,
\hspace{5mm}
\overline{\cal O}_{i}=\frac{{\cal
O}_{i}}{\sqrt{\langle D_{i}\rangle}}~,\hspace{5mm} 
\langle
D_{i}\rangle\equiv\langle\widetilde{N,0}|D_{i}|\widetilde{N,0}\rangle~,\hspace{5mm}
i=p,~h~,
\label{Obar}
\end{equation}
where the correlated ground state 
$|\widetilde{N,0}\rangle$ is defined as the vacuum of the operators ${\cal
A}_{\mu}$ and ${\cal
R}_{\lambda}$, i.e.
\begin{equation}
    {\cal A}_{\mu}|\widetilde{N,0}\rangle={\cal
    R}_{\lambda}|\widetilde{N,0}\rangle=0~.
    \label{GS}
    \end{equation}
The GSC factors $\langle D_{i}\rangle$ are the expectation values of the
operators
\begin{equation}
    D_{p}=1-N_{p}~,\hspace{5mm} D_{h}=N_{h}-1~,
    \label{D}
\end{equation}
in the correlated ground state $|\widetilde{N,0}\rangle$ (\ref{GS}). They 
are introduced in the definition (\ref{calAR}) to
preserve the ground-state expectation value of the commutation relation (\ref{[PP]}).
Indeed, using Eq. (\ref{[PP]}) and definition (\ref{Q}), one obtains
the exact commutation relations for operators $Q_{i}$ and
$Q^{\dagger}_{i}$ in the form
\begin{equation}
[Q_{p}, Q_{p'}^{\dagger}]=\delta_{pp'}D_{p}~,\hspace{5mm} 
[Q_{h}, Q_{h'}^{\dagger}]=\delta_{hh'}D_{h}~.
\label{[QQ]}
\end{equation}
Using this exact relation (\ref{[QQ]}) and the definition 
(\ref{GS}), one can see that the renormalized addition and removal operators satisfy the
boson commutation relations in the correlated ground state (\ref{GS})
\begin{equation}  
    \langle[{\cal A}_{\mu},{\cal A}_{\mu'}^{\dagger}]\rangle=\delta_{\mu\mu'}~,\hspace{5mm} 
    \langle[{\cal R}_{\lambda},{\cal R}_{\lambda'}^{\dagger}]\rangle=\delta_{\lambda\lambda'}~,
    \label{R[AA][RR]}
\end{equation}
if the amplitudes ${X}$ and ${Y}$ satisfy the same 
RPA orthogonality conditions (\ref{norm}), while the same RPA closure 
relations (\ref{closure}) 
guarantee 
the following inverse transformation of Eq. (\ref{calAR}) 
\[
Q_{p}^{\dagger}=\sqrt{\langle 
D_{p}\rangle}\bigg[\sum_{\mu}{X}_{p}^{(\mu)}{\cal A}_{\mu}^{\dagger}
+\sum_{\lambda}{Y}_{p}^{(\lambda)}{\cal R}_{\lambda}\bigg]~,
\]
\begin{equation}
Q_{h}=\sqrt{\langle 
D_{h}\rangle}\bigg[\sum_{\lambda}{
X}_{h}^{(\lambda)}{\cal R}_{\lambda}
+\sum_{\mu}{ Y}_{h}^{(\mu)}{\cal A}_{\mu}^{\dagger}\bigg]~.
\label{inverseR}
\end{equation}
instead of Eq. (\ref{inverse}).

The RRPA equations are obtained in the same way as that for the
derivation of the RPA equations, assuming the factorization 
\begin{equation}
    \langle D_{i}D_{j}\rangle\simeq\langle D_{i}\rangle\langle 
    D_{j}\rangle~,
    \label{approx}
\end{equation}
as well as neglecting all the expectation values 
$\langle Q^{\dagger}_{p'}Q_{p}\rangle$, $\langle Q_{p}Q_{h}\rangle$, 
and $\langle Q_{h}^{\dagger}Q_{h'}\rangle$. The RRPA submatrices
obtained in this way for the addition modes have the form
\begin{equation}
 A_{pp'}^{\rm RRPA}=
 2\bigg[\epsilon\bigg(p-\frac{1}{2}\bigg)+\frac{G}{2}\bigg]
 \delta_{pp'}
-G\sqrt{\langle D_{p}\rangle\langle D_{p'}\rangle}~,
\label{ARRPA}
\end{equation}
\begin{equation}
B_{ph}^{\rm RRPA}=
G\sqrt{\langle D_{p}\rangle\langle 
D_{h}\rangle}~,
\label{BRRPA}
\end{equation}
\begin{equation}
 C_{hh'}^{\rm RRPA}= 
 -2\bigg[\epsilon\bigg(h-\frac{1}{2}\bigg)+\frac{G}{2}\bigg]\delta_{hh'}
+G\sqrt{\langle D_{h}\rangle\langle D_{h'}\rangle}~.
\label{CRRPA}
\end{equation}
They are different from the RPA submatrices (\ref{ARPA}) -- (\ref{CRPA})
by the square roots $\sqrt{\langle D_{i}\rangle\langle D_{j}\rangle}$, which renormalize the
interaction. In the absence of GSC beyond RPA the expectation values $\langle
D_{i}\rangle$
become 1, and the RPA submatrices are recovered.

The equations for the expectation values $\langle D_{i}\rangle$ are obtained
following Refs. \cite{CaDaSa,Duk,Dang} in the form
\begin{equation}
    \langle D_{p}\rangle=\frac{1}{1+2\sum_{\lambda}[Y_{p}^{(\lambda)}]^{2}}~,
    \hspace{5mm} 
    \langle D_{h}\rangle=\frac{1}{1+2\sum_{\mu}[Y_{h}^{(\mu)}]^{2}}~,
    \label{DpDh}
\end{equation}
which are exact in the present model because of the exact
relation
\begin{equation}
    N_{i}=2P_{i}^{\dagger}P_{i}~.
\label{N}
\end{equation}
For the details see Ref. \cite{Duk}.
The set of Eqs. (\ref{RPA}), (\ref{ARRPA}) -- (\ref{CRRPA}), and
(\ref{DpDh}) forms the closed set
of RRPA equations, which are non-linear due to the presence of the 
backward-going amplitudes $Y$ in the expressions for the GSC factors $\langle
D_{i}\rangle$ (\ref{DpDh}). In numerical calculations this set is solved self-consistently by
iteration. Knowing $\langle D_{i}\rangle$, one obtains 
the occupation numbers $f_{i}$ from Eq. (\ref{D}) as
 \begin{equation}
     f_{p}=\frac{1}{2}(1-\langle D_{p}\rangle)~,\hspace{5mm}
     f_{h}=\frac{1}{2}(1+\langle D_{h}\rangle)~,
     \label{fi}
     \end{equation}
     so that within the RPA ($D_{i}=$ 1) the HF
     occupation numbers $f_{p}=$ 0 and $f_{h}=$ 1 are recovered.
\subsection{SCRPA}
The only difference between the RRPA and SCRPA is that the latter
includes the so-called screening factors, which are the expectations values
$\langle Q^{\dagger}_{p'}Q_{p}\rangle$, $\langle Q_{p}Q_{h}\rangle$, 
and $\langle Q_{h}^{\dagger}Q_{h'}\rangle$ neglected within the RRPA. 
They are derived using the inverse transformation (\ref{inverseR}) 
and the definition of the correlated ground state (\ref{GS}), and have the
form~\cite{Duk,Dang}:
\begin{equation}
\langle Q_{p}^{\dagger}Q_{p'}\rangle=\langle P_{p}^{\dagger}P_{p'}\rangle=
\sqrt{\langle D_{p}\rangle\langle 
D_{p'}\rangle}\sum_{\lambda}Y_{p}^{(\lambda)}Y_{p'}^{(\lambda)}~,
\label{pp}
\end{equation}
\begin{equation}
\langle Q_{p}Q_{h}\rangle=\langle Q_{h}^{\dagger}Q_{p}^{\dagger}\rangle=
-\langle P_{h}^{\dagger}P_{p}\rangle=
-\langle P_{p}^{\dagger}P_{h}\rangle=
\sqrt{\langle D_{p}\rangle\langle 
D_{h}\rangle}\sum_{\lambda}X_{h}^{(\lambda)}Y_{p}^{(\lambda)}~,
\label{ph}
\end{equation}
\begin{equation}
\langle Q_{h}^{\dagger}Q_{h'}\rangle=
\sqrt{\langle D_{h}\rangle\langle 
D_{h'}\rangle}\sum_{\mu}Y_{h}^{(\mu)}Y_{h'}^{(\mu)}=
\langle P_{h'}^{\dagger}P_{h}\rangle
-\delta_{hh'}\langle 
D_{h}\rangle~,
\label{hh}
\end{equation}
\[
{\rm with}\hspace{3mm} 
\langle P_{h'}^{\dagger}P_{h}\rangle=\sqrt{\langle D_{h}\rangle\langle 
D_{h'}\rangle}\sum_{\lambda}X_{h}^{(\lambda)}X_{h'}^{(\lambda)}~.
\]

The SCRPA submatrices are then given as~\cite{Duk,Dang}
\[
A_{pp'}=\langle[~\overline{Q}_{p},
[H,\overline{Q}_{p'}^{\dagger}]]\rangle=
\]
\begin{equation}
    2\bigg\{\bigg[\epsilon\bigg(p-\frac{1}{2}\bigg)+\frac{G}{2}\bigg]+
\frac{G}{\langle D_{p}\rangle}\bigg[\sum_{p''}\langle 
Q_{p''}^{\dagger}Q_{p}\rangle-\sum_{h''}\langle 
Q_{p}Q_{h''}\rangle\bigg]\bigg\}\delta_{pp'}
-G\frac{\langle D_{p}D_{p'}\rangle}{\sqrt{\langle D_{p}\rangle\langle 
D_{p'}\rangle}}~,
\label{Amatrix}
\end{equation}
\begin{equation}
B_{ph}=-\langle[~\overline{Q}_{p},
[H,\overline{Q}_{h}]]\rangle=
G\frac{\langle D_{p}D_{h}\rangle}{\sqrt{\langle D_{p}\rangle\langle 
D_{h}\rangle}}~,
\label{Bmatrix}
\end{equation}
\[
C_{hh'}=-\langle[~\overline{Q}_{h},
[H,\overline{Q}_{h'}^{\dagger}]]\rangle=
\]
\begin{equation}
    -2\bigg\{\bigg[\epsilon\bigg(h-\frac{1}{2}\bigg)+\frac{G}{2}\bigg]+
\frac{G}{\langle D_{h}\rangle}\bigg[\sum_{h''}\langle 
Q_{h}^{\dagger}Q_{h''}\rangle-\sum_{p''}\langle 
Q_{p''}^{\dagger}Q_{h}^{\dagger}\rangle\bigg]\bigg\}\delta_{hh'}
+G\frac{\langle D_{h}D_{h'}\rangle}{\sqrt{\langle D_{h}\rangle\langle 
D_{h'}\rangle}}~.
\label{Cmatrix}
\end{equation}
In principle no assumption on the factorization (\ref{approx}) is
required within the SCRPA. However the factorization (\ref{approx})
does simplify greatly the calculations, whose results turn out to be
quite close to those obtained without assuming
(\ref{approx})~\cite{Duk}. The
equations for the GSC factors are the same as Eq. (\ref{DpDh}).
\subsection{Two-particle separation energy and pair transition}
\subsubsection{Consistency in terms of two-particle separation energy}
\label{cons}
Unlike the particle-hole RPA, where the solutions with negative
eigenvalues have no physical meaning, each set of the
(particle-particle) RPA, RRPA, and SCRPA equations has $m+n$ 
solutions, all of which correspond to physical states. 
From them $m$ ($=N/2$ in the present case) solutions correspond to the system with $N+2$
particles, whose excitation 
energies related to the ground-state (g.s.) energy
${\cal E}_{\rm g.s.}^{(N)}$ 
of the N-particle system are 
${E}_{\mu}\equiv{\cal E}_{\mu}^{(N+2)}-{\cal E}_{\rm g.s.}^{(N)}$ 
with $\mu
=1,\ldots, m$. The 
other $n$ $(=N/2)$ solutions have the eigenvalues equal to  
$-{E}_{\lambda}\equiv-[{\cal E}_{\rm g.s.}^{(N)}-{\cal
E}_{\lambda}^{(N-2)}]$ with $\lambda=1,\ldots, n$. They 
correspond to the eigenstates of the system with $N-2$ particles with
the eigenvalues ${E}_{\lambda}$. Therefore the 
first eigenvalues $E_{\mu=1}$ of the $(N+2)$-particle system 
and $E_{\lambda=1}$ of the $(N-2)$-particle one are
the ground-state energies of these systems related to 
that of the $N$-particle core.
This means that they correspond to the two-particle
separation energies ${S}_{2p}^{(+)}(N+2)$ in the $(N+2)$-particle system,
and ${S}_{2p}^{(-)}(N)$ in the core, which are defined as
\begin{equation}
    {S}_{2p}^{(+)}(N+2)={E}_{\mu=1}\equiv{\cal E}_{\rm g.s.}^{(N+2)}-{\cal E}_{\rm
    g.s.}^{(N)}~,\hspace{5mm} 
    {S}_{2p}^{(-)}(N)={E}_{\lambda=1}\equiv{\cal E}_{\rm g.s.}^{(N)}-{\cal E}_{\rm
    g.s.}^{(N-2)}~.
\label{S2n}
\end{equation}
As such, in an exact or consistent theory, it follows from Eq.
(\ref{S2n}) that
\begin{equation}
    S_{2p}^{(+)}(N+2)=S_{2p}^{(-)}(N+2)~.
    \label{equal}
    \end{equation}
This equation is called the consistency condition hereafter, which
means that the two ways of calculating the two-particle separation
energy as the energy of the addition mode, which adds two particles
to the $N$-particle core, and that of the removal mode, which removes 
two particles from the $(N+2)$-particle system to get back to the same
core, should give the same result.
The expressions (\ref{S2n}) were used in Ref. \cite{VinhMau} to calculate the
two-neutron separation energies for $N_{c}+2$ and $N_{c}$ nuclei, 
namely $^{12,14}$C,
$^{10,12}$Be, and $^{9,11}$Li, whose neutron cores
consist of $N_{c}=$ 8 particles. 
\subsubsection{Pair-transition probability}
    The pair transition is generated by the operator
    \begin{equation}
	K=\sum_{i=p,h}(P_{i}^{\dagger}+P_{i})=\sum_{p}(Q_{p}^{\dagger}
	+Q_{p}) - \sum_{h}(Q_{h}^{\dagger}
	+Q_{h})~.
	\label{K}
	\end{equation}
	By using the inverse transformation (\ref{inverseR}), one
	obtains the square $|\langle\widetilde{N,0}|{\cal
	A}_{\mu}K|\widetilde{N,0}\rangle|^{2}$ 
	of the matrix element of the operator $K$,
	which corresponds to the
    transition probability from the
    ground state of the $N$-particle system to the one-phonon state $|\mu\rangle$ of the $(N+2)$-particle 
    system within the RRPA and/or SCRPA, in the following form
    \begin{equation}
	B_{\mu}(N \rightarrow N+2)\equiv|\langle
	\widetilde{N,0}|{\cal A}_{\mu}K|\widetilde{N,0}\rangle|^{2}=\bigg|\sum_{p}\sqrt{\langle
	D_{p}\rangle}X_{p}^{(\mu)}-\sum_{h}\sqrt{\langle
	D_{h}\rangle}Y_{h}^{(\mu)}\bigg|^{2}~.
	\label{Bmu}
	\end{equation}
	In the same way, the transition probability from the 
	ground state of the $N$-particle system to the one-phonon state $|\lambda\rangle$
	of the $(N-2)$-particle system is proportional to $|\langle
	\widetilde{N,0}|{\cal R}_{\lambda}K|\widetilde{N,0}\rangle|^{2}$,
	which is given as
	\begin{equation}
	       B_{\lambda}(N-2\leftarrow
	       N)\equiv|\langle
	       \widetilde{N,0}|{\cal R}_{\lambda}K|\widetilde{N,0}\rangle|^{2}=
	       \bigg|\sum_{h}\sqrt{\langle
	       D_{h}\rangle}X_{h}^{(\lambda)}-\sum_{p}\sqrt{\langle
	       D_{p}\rangle}Y_{p}^{(\lambda)}\bigg|^{2}~.
	       \label{Blambda}
	       \end{equation}
	       The corresponding expressions within the RPA are
	       recovered by setting $\langle D_{i}\rangle=$ 1 in Eqs. 
	       (\ref{Bmu}) and (\ref{Blambda}). From the 
	       consistency condition discussed in the preceding
	       section it follows that
	       \begin{equation}
		   B_{\mu=1}(N \rightarrow N+2)=  B_{\lambda=1}(N\leftarrow
	       N+2)~,
	       \label{prob}
	       \end{equation}
	       which means that the probabilities of the pair
	       transition
	       between the ground states of $N$- and $(N+2)$-particle
	       systems should be the same in either direction.  
	       
	       In the limit with
	       $\epsilon_{p>1}=\epsilon_{p=1}=\epsilon/2$ and 
	       $\epsilon_{h>1}=\epsilon_{h=1}=-\epsilon/2$, 
	       the Richardson model is reduced to the well-known two-level
	       model, in which each 
of the two levels is $N$-fold and the Fermi level (\ref{lambda})
is located at $\epsilon_{\rm F}=-G/2$. 
This degenerate model has been
studied in detail in Refs. \cite{Hoga,Hagino,Dang1}. The 
RRPA solutions of this model yields the following two-particle separation
energies [See, e.g, Eqs. (73) and (74) of Ref. \cite{Dang1}] 
\begin{equation}
S_{2p}^{(+)}(N+2)=\sqrt{(\epsilon+G)[\epsilon-G(\langle D \rangle N-1)]}~,
\hspace{5mm} 
S_{2p}^{(-)}(N+2)=\sqrt{(\epsilon+G)[\epsilon-G(\langle D\rangle N +1)]}~.
\label{S}
\end{equation}
The RPA result is recovered by setting $\langle D\rangle =$ 1. 
Result (\ref{S}) analytically shows that, for this degenerate two-level model, 
the consistency condition (\ref{equal}) is 
asymptotically fulfilled only when $N\gg$ 1.
\section{Analysis of numerical calculations}
\label{numerical}
The calculations were carried out within the Richardson model with the
level distance $\epsilon=$ 1 MeV at several values of the particle
number $N$ as the pairing parameter $G$ varies.

\begin{figure}                                                             
\includegraphics[width=12cm]{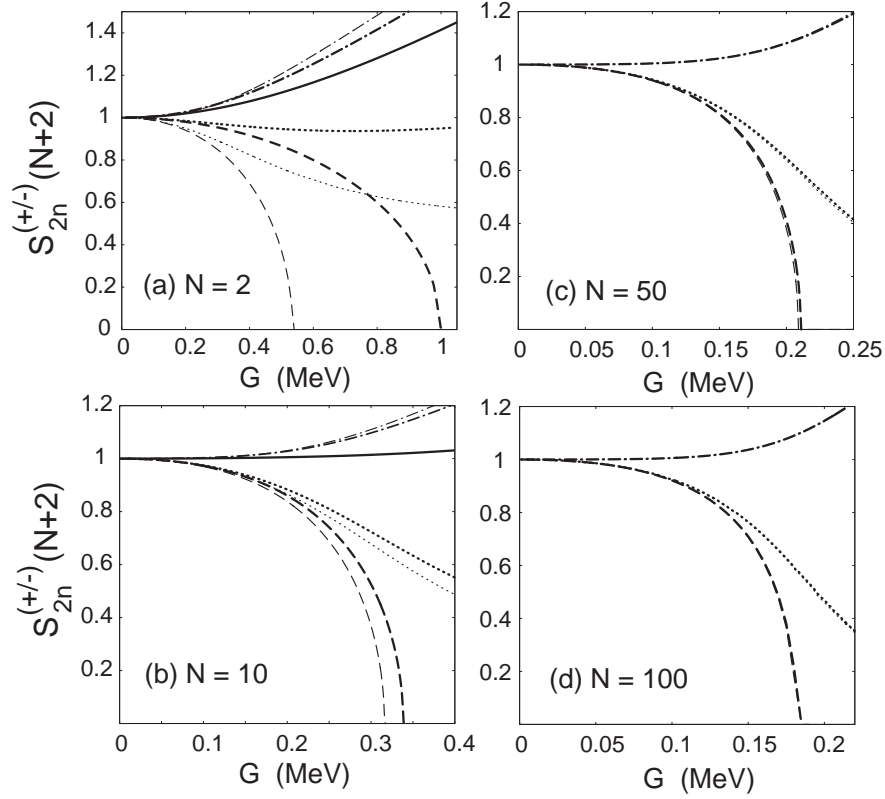}
\caption{\label{energy}
Two-particle separation energies
$S_{2n}^{(\pm)}(N+2)$ at several
values of $N$ as functions of pairing parameter $G$. 
The dashed, dotted, and dash-dotted lines
represent results obtained within RPA, RRPA, and SCRPA, respectively, 
for which the thin lines denote $S_{2n}^{(-)}(N+2)$, 
while the thick lines stand for $S_{2n}^{(+)}(N+2)$. 
The exact results for $N=$ 2 and 10 are shown by the thick solid lines.}
\end{figure}
\begin{figure}                                                             
\includegraphics[width=12cm]{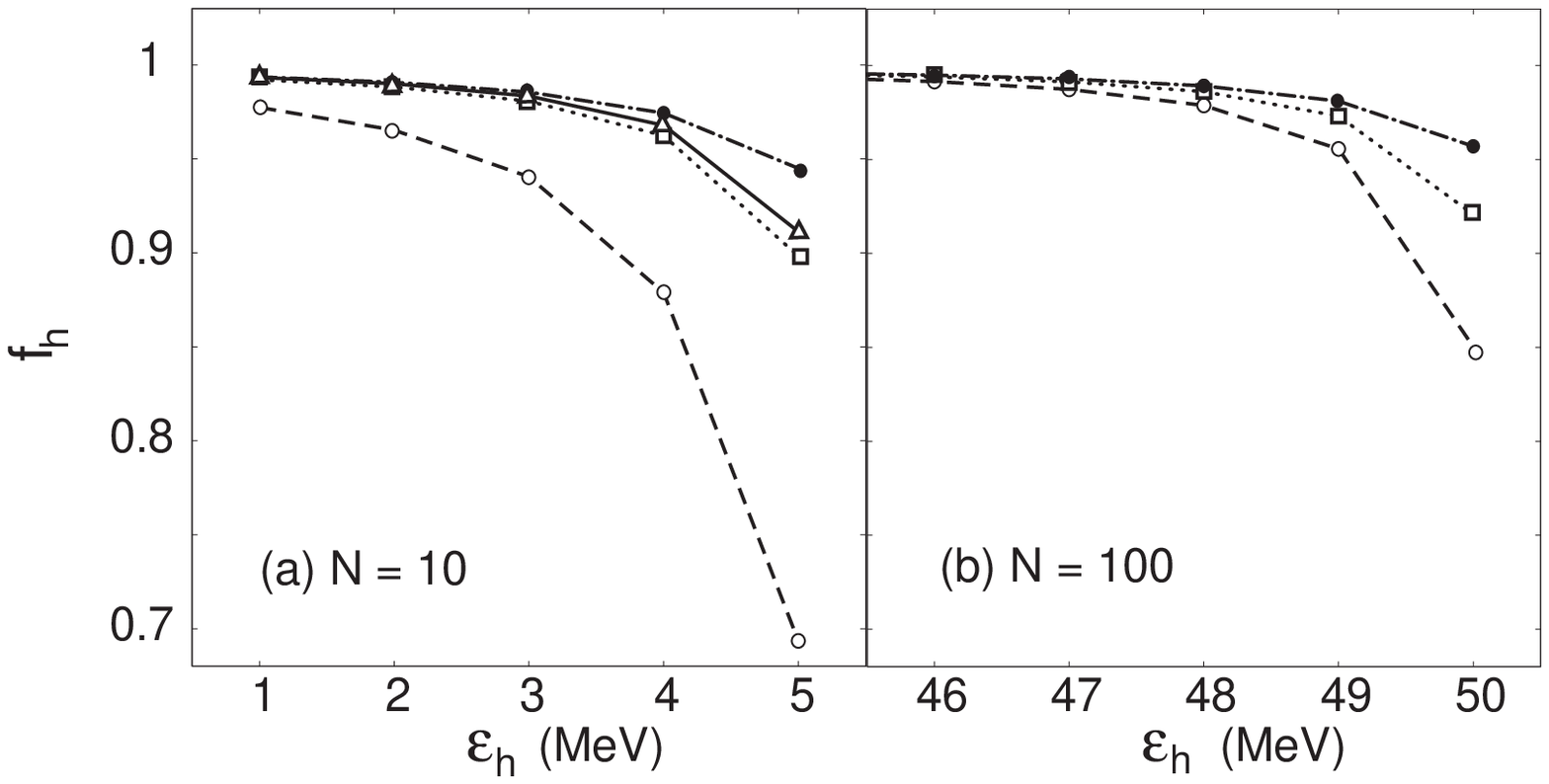}
\caption{\label{fh}
Occupation numbers $f_{h}$ of five hole levels nearest to the
Fermi level ($h=$ 1,\ldots, 5) for the systems with $N=$ 10 (a) and 100 (b).  
The open circles, squares, and full circles denote the
RPA, RRPA, and SCRPA results, respectively. The triangles in (a) are
the exact results. The dashed, dotted, 
dash-dotted, and solid lines connecting these discrete points are drawn to guide the
eye. The results
in (a) and (b) are obtained by using $G=$ 0.33 MeV and 0.17 MeV,
respectively.}
\end{figure}
Shown in Fig. \ref{energy} are the two-particle separation energies
$S_{2n}^{(\pm)}(N+2)$, which are obtained
within the RPA, RRPA, and SCRPA according to Eq. (\ref{S2n}) at $N=$
2, 10, 50, and 100. For a comparison, the exact results,
available for $N=$ 2 and 10, are also shown. For $N=$ 50 and 100
the size of the matrix makes the exact diagonalization  
infeasible so that the exact solutions are not available. In Ref. \cite{Duk} it has
been analyzed in detail that the RRPA and SCRPA extend the solutions
far beyond the point where the RPA collapses, and that the SCRPA
predictions are the closest ones to the exact solutions. Therefore
the details of these features are not repeated here. This figure shows 
that, for $N\leq$ 10, the consistency condition (\ref{equal}) is
clearly violated in all the approximations under consideration. The violation is
particular strong within the RPA at small
$N$. It is seen that
$S_{2n}^{(+)}(N+2) > S_{2n}^{(-)}(N+2)$ within the RPA and RRPA, but 
$S_{2n}^{(+)}(N+2) < S_{2n}^{(-)}(N+2)$ within the SCRPA. 
At a given value of $G$, the SCRPA predicts the smallest 
difference between $S_{2n}^{(+)}(N+2)$
and $S_{2n}^{(-)}(N+2)$, while the RPA gives the largest one. 
One can see how 
$S_{2n}^{(+)}(N+2)$ gets closer to $S_{2n}^{(-)}(N+2)$ as $N$
increases so that Eq. (\ref{equal}) is perfectly restored at $N=$
100. The small difference between $S_{2n}^{(+)}(N+2)$
and $S_{2n}^{(-)}(N+2)$ and the better agreement with the exact
result, obtained within the SCRPA even at small $N$, show that the SCRPA is the
most consistent approximation among three approximations under
consideration. At the
same time, the quite large difference between $S_{2n}^{(+)}(N+2)$
and $S_{2n}^{(-)}(N+2)$ obtained within the RPA at small $N$ 
shows that the RPA is, in principle, an inconsistent theory when
applied to light systems. This observation means that, GSC 
are important and cannot be neglected in calculations for
light systems.

\begin{figure}                                                             
\includegraphics[width=12cm]{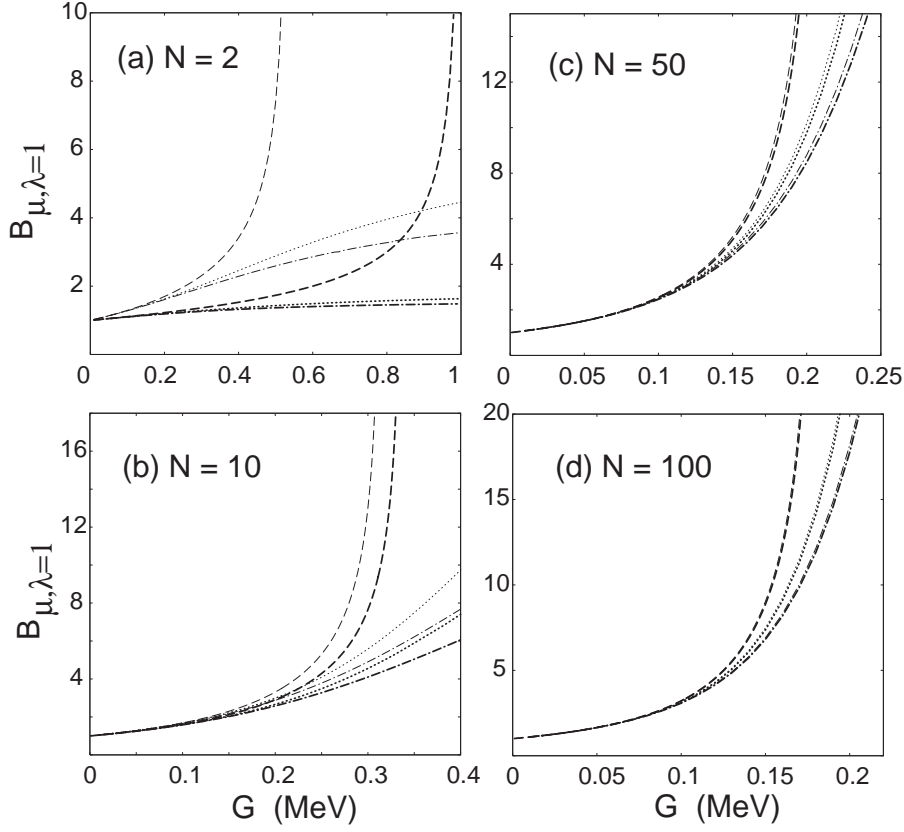}
\caption{\label{BE}
Squares of matrix elements corresponding to the pair-transition probabilities 
as functions of pairing parameter $G$ at
several values of $N$.
The notations for dashed, dotted, and dash-dotted lines are as in Fig.
\ref{energy}. The thick and thin lines denote $B_{\mu=1}(N\rightarrow N+2)$
and $B_{\lambda=1}(N\leftarrow N+2)$, respectively.}
\end{figure}
In order to see the origin of the 
effect due to GSC, the occupation numbers $f_{h}$ of five hole
levels, which are located nearest to the Fermi level, are plotted in Fig. \ref{fh} 
for $N=$ 10 and 100. They are obtained by using the values of $G$
equal to 0.33 MeV and 0.17 MeV for $N=$ 10 and
100, respectively. These values are just below 
the collapsing points of the RPA, which are $G_{\rm crit}=$ 0.34 MeV and 0.18 MeV
for $N=$ 10 and 100, respectively. The results in Fig. \ref{fh} (a) shows that the
QBA, on which the RPA is based, is no longer a good approximation when $N$ is small. Indeed, the
QBA means that $f_{h}=$ 1, while the value of $f_{h}$, obtained for the
level just below the Fermi level, is around 0.7 within the RPA.
At large $N$, e.g. $N=$ 100, the QBA is much better fulfilled as the 
occupation number $f_{h=1}$ is around 0.85,
while those for all the other 49 hole levels are larger than 0.95 [Fig. \ref{fh}
(b)]. This figure also shows that the effect of GCS, once included
within the RRPA and SCRPA, makes the RRPA and SCRPA phonon operators much
closer to the ideal bosons since the values of $f_{h}$ predicted by the
RRPA and SCRPA for the hole level just below the Fermi level 
are around 0.9 and 0.95, respectively, for $N=$ 10, and larger than
these values at larger $N$. The comparison of
the occupation numbers predicted by 
the RPA, RRPA, and SCRPA again shows that the QBA is best satisfied
within the SCRPA. It is interesting to notice that, while the 
SCRPA predicts a two-particle separation energy much closer
to the exact value for $N=$ 10 as compared to the RPA [Fig.
\ref{energy} (b)], the agreement between the SCRPA prediction and exact
values for the occupation number $f_{h=1}$ is not as good as that
offered by the RRPA. 

The pair-transition probabilities, shown in Fig. \ref{BE}, increase with
increasing $G$ as one might expect. In a similar way as that for 
the two-particle separation energies, the condition (\ref{prob}) is
strongly violated within the RPA for small $N$. Combining the results on 
Figs. \ref{energy} and \ref{BE}, it is seen that for $N\leq$ 10, 
neither condition (\ref{equal}) nor (\ref{prob}) holds. Like the
restoration of the consistency condition (\ref{equal}) for two-particle separation
energies, as $N$ increases
the condition (\ref{prob}) for pair-transition probabilities 
is also gradually restored to become well-fulfilled at $N=$ 100.
\section{Conclusions}
The present paper tested the consistency of the RPA,
RRPA, and SCRPA within the Richardson model.
The consistency condition under consideration here is the requirement 
that the two-particle separation energy $S_{2n}^{(+)}(N+2)$ obtained
as the energy of the
first addition mode, which adds 2 particles to the $N$-particle
system, should be the same as $S_{2n}^{(-)}(N+2)$ obtained as the energy of the first
removal mode, which removes 2 particles from the $N+2$-particle
system. The results of calculations show that this consistency
condition is well
fulfilled only for sufficiently large values of the particle number
$N$ ($\geq$ 50 in the present model), i.e. in medium-mass and heavy systems.
For light systems ($N\leq$ 10) it is found that, among the three approximations under consideration, 
the RPA, which has the largest deviation from the exact
results at large $G$, strongly violates the consistency condition.
The SCRPA agrees best with the exact solutions and also is the most
consistent approximation in the sense that the above-mentioned
consistency condition is well satisfied already at $N\geq$ 10 even at 
$G$ above the RPA collapsing point.

The results obtained in this test indicate that GSC beyond RPA becomes 
quite important in light systems, especially in the region close to the
RPA collapsing point. They invalidate the QBA, which is assumed 
for the pair operators within the RPA. Therefore, in principle, the RPA predictions of
the two-neutron separation energies in light neutron-rich nuclei
cannot be reliable without taking the effect of GSC beyond RPA into account. From the
three approximations considered in the present paper, the most
consistent one, which includes this effect for light systems, turns out to
be the SCRPA. Although the present test was conducted within a schematic model with 
doubly-fold equidistant levels, it is unlikely that this general conclusion on the
inconsistency of the RPA would change when applied to realistic light 
nuclei. As a matter of fact, the recent calculations of the
two-neutron separation energy $S_{2n}$ in $^{12,14}$Be by using the Gogny
interaction have shown that the RRPA shifts the result by around 14$\%$
for $^{12}$Be, and the consistency condition
(\ref{equal}) is not fulfilled ($N=$ 10) for both the RPA and RRPA~\cite{DangVM}. 
An easy way to see if the RPA needs to be renormalized 
is to estimate the occupation numbers
(\ref{fi}) using the backward-going amplitudes $Y$ obtained within the RPA. 
If there is a strong deviation of $f_{p}$ ($f_{h}$) from 0
(1), this means that the validity of the RPA is deteriorated.
In the above-mentioned calculations for Be-isotopes e.g., 
the occupation number of the last occupied shell amounts to around 0.8 within the RPA, and
increases to around 0.95 within the RRPA, showing the necessity of
using the SCRPA, or at least the RRPA, instead of the RPA. 
\acknowledgments
The numerical calculations were carried out using the {\scriptsize 
FORTRAN IMSL} Library by Visual Numerics on the RIKEN Super 
Combined Cluster (RSCC) System. Thanks are due 
to Nicole Vinh Mau (Orsay) for fruitful discussions, and Alexander
Volya (Tallahassee) for the exact solutions of the Richardson model.

\end{document}